# From error detection to behaviour observation: first results from screen capture analysis

*Françoise Tort, François-Marie Blondel, Éric Bruillard*
*UMR STEF – INRP – ENS Cachan, UniverSud*
*61, avenue du Président Wilson*
*94235 Cachan cedex - France*
*francoise.tort@ecogest.ens-cachan.fr*
*francois-marie.blondel@inrp.fr*
*eric.bruillard@creteil.iufm.fr*

**ABSTRACT**

*This paper deals with errors in using spreadsheets and analysis of automatic recording of user interaction with spreadsheets. After a review of literature devoted to spreadsheet errors, we advocate the importance of going from error detection to interaction behaviour analysis. We explain how we analyze screen captures and give the main results we have obtained using this specific methodology with secondary school students (N=24). Transcription provides general characteristics: time, sequence of performed tasks, unsuccessful attempts and user preferences. Analysis reveals preferred modes of actions (toolbar buttons or menu commands), ways of writing formulas, and typical approaches in searching for solutions. Time, rhythm and density appear to be promising indicators. We think such an approach (to analyze screen captures) could be used with more advanced spreadsheet users.*

## 1. INTRODUCTION

Several studies have been devoted to the detection of errors and mistakes in spreadsheets, leading to some agreement concerning typology of errors (Panko & Halverson 1996; Chadwick & Sue 2001, Panko 2008). These studies have great importance for formulating strategies for identifying possible errors in spreadsheets. Are they completely conclusive?

One can imagine that errors are classical and stable in the sense that users in the eighties and "new millennium" users are expected to commit the same ones. But, these "new users" may have new ways of doing things, have acquired new routines leading to different kinds or patterns of errors. Adopting an educational point of view also leads to go beyond errors, notably to discover their origin.

During the DidaTab project (Blondel et al. 2008), we designed paper and computer tests to assess the competency levels of students. Observations made during these tests revealed that the way students work with spreadsheet software is as interesting as the final result they obtain. Their way of doing so is not always simple and is often surprising. Starting from these observations, we try to consider errors mainly as symptoms, inviting us to better understand the process underlying mistakes.

Furthermore it is useful to distinguish competency and performance. Not to be able to solve a particular question often is a weak indicator of understanding. The question is not only what students' answers are but how they proceed, leading to other issues. Can we infer from the way they respond to tests what their level of understanding of spreadsheet objects is? Do tested students show characteristic approaches and behaviours allowing us to better understand their way of working with software such as spreadsheets? What difficulties do they encounter?

Little empirical data are available to researchers to analyze the processes and actions of end-users.

In the first part of this paper, we give an overview of classic literature on spreadsheet errors and advocate the necessity to go more in-depth in analyzing the solving processes. In the second part, we describe different methodological approaches to get an outline of the user interaction process using spreadsheets. In the third part, we present the first results obtained in an experiment of screen capture analysis of students working with spreadsheets. Finally we propose some discussion.

## 2. SPREADSHEET ERRORS AND LEARNING PROCESSES

Panko and Halverson (1996) revisited by Panko (2008) offer a very complete taxonomy of spreadsheet errors. They distinguish between *quantitative* and *qualitative* errors. Quantitative errors lead to incorrect final values. Qualitative errors refer to risky practices that could lead to errors in later uses of the spreadsheet. The authors further divided the quantitative category, into mechanical errors, logic errors and omission errors. The classification criteria are relatively simple, and provide a reliable classification.

Rajalingham et al. (2000) propose a more complex taxonomy of spreadsheet errors. They make the same *quantitative* and *qualitative* distinction as Panko and Halverson. They distinguish *accidental* errors from *reasoning* errors (similar to mechanical errors and logic errors) and introduce an interesting difference between *developer* errors made during the development of the spreadsheet and *end-user* errors, covering errors made by users manipulating a spreadsheet after its creation.

Taxonomies allow the comparison of errors across studies and application areas. They are important to formulate strategies for discovering possible errors in operational spreadsheets, in order to fix them or prescribe a kind of cure.

In the educational context of students learning spreadsheets, we study abilities and difficulties of, and obstacles to, spreadsheet learners. Errors are mainly considered as symptoms and indicators of particular difficulties. Taxonomies can be used to diagnose the cause of errors. But the main point is not to detect errors and fix them, but to know how errors are created and what their causes are. Therefore, we are interested not only in students' spreadsheet errors but also in the process of their creation.

### 2.1. Students' spreadsheet errors

We will focus on two of the studies which have analysed students' spreadsheet abilities, dealing with the possible origins of errors.

In an experimental study comparing the traditional spreadsheet calculation paradigm with a structured spreadsheet calculation paradigm, Tukiainen (2000) tried to explain errors made by novice users in terms of the computation "*doctrine*" (spreadsheet-based programming as opposed to other programming techniques), and of the spreadsheet calculation tool's usability issues.

Reinhardt and Pillay (2005) aimed to develop a spreadsheet tutor helping students to overcome their learning difficulties. They studied what students were doing during a spreadsheet test in order to build an error library and a model to provide individualized feedback. Their own classification focuses on errors in formulae and functions, and distinguishes between three errors categories, depending on their level of generality: errors that apply to any function and formula, errors that occur in formulae and functions with a similar syntax, errors that only occur in specific formulae or functions. They provide interesting and detailed examples.

In the framework of the DidaTab project, we administered computer-based tests to about 200 French secondary school students (Blondel et al., 2008). Students had to perform very basic tasks: change cells format, write a formula (example in figure 1), create a graphic or sort data (see Tort and Blondel 2007 for methodological details).

**Figure 1:** Task 3, Write the sum and the average of a vector

Our results revealed spreadsheet errors already detailed in (Tukiainen, 2000) or (Reinhardt & Pillay, 2005) studies. Some errors we found are not described in the literature and appear more difficult to classify in known taxonomies. Most of them were made by initial spreadsheet users (Table 1).

1. Plotting a series of data available on a line, including the total amount which is displayed at the end of the line.
2. Sorting only one column of data instead of the whole table of data.
3. Erasing digits after the decimal point of a number in a cell, in order to display only one digit after the decimal point, instead of changing the display format of the cell.
4. Using functions in non-conventional manners like SUM(A1+A2+A3) or PRODUCT(A1*12,5) or SUM(A1*12,5).
5. Writing successive formulae with relative cell references, instead of writing one formula with an absolute cell reference and copying it with the fill handle.

**Table 1:** Some errors of initial spreadsheet users observed during DidaTab project

Some errors can be considered as general errors. For instance, example 1 is the transposition in the context of the creation of a graphic of the error GE4c *"Additional cells included in argument. Student included the title or additional cells at the end or at the beginning of the necessary block of cells."* described in (Reinhardt & Pillay, 2005)

According to the Panko and Halverson (1997) classification, examples 1 and 2 are quantitative errors. They could be logical errors, if we assume that students applied a wrong method. They could be mechanical errors, if students obtained such bad results because they agreed to a suggestion automatically proposed by the spreadsheet software. Example 3 is a pure quantitative logical error. Examples 4 and 5 are qualitative errors, not visible from displayed values, but risky for later use of the spreadsheet.

All these errors are of high interest, but their origin is difficult to infer: traces of confusion of students while using the spreadsheet software, symptoms of students misunderstanding of spreadsheet objects and principles, or both.

Our aim is to investigate how these errors have been created and why.

### 2.2. Errors creation process of initial spreadsheet users

Powell et al. (2005) point out the fact that relatively little is known about the creation of spreadsheet errors. They notice that only a few laboratory experiments have been carried out, and no one has attempted to study spreadsheet development in the field, at a level of detail allowing observation of developers making errors.

Most of the students we have tested were relative novices in the use of spreadsheets, but had already personal experience of ICT. We can assume that these "new learners" have developed their own ways of working with new digital objects and computer applications. Chadwick and Sue (2001) observe that novices overrate their own literacy skills and make errors despite being quite confident. It is likely that they are satisfied most of the time by their own ways of doing things. We presume that their prior experience with other software is involved in the process of spreadsheet errors creation.

In the following section, we propose a methodological framework to observe learners creating spreadsheets at a very high level of detail.

## 3. METHODOLOGICAL APPROACHES

### 3.1. Data collection for process analysis

Creating spreadsheets may be considered as a particular case of software development. Human-Computer Interaction researchers have analysed user behaviour during software development with the intention of understanding what cognitive skills are involved.

In a paper on the various analysis that can be used to study the way users interact with a computer applications, Hulshof (2004) discusses the most commonly used methods to track interaction: eye tracking, thinking aloud and log files. Eye tracking provides researchers with precise information about the type of information users are processing, but interpretation may be difficult when high level cognitive processes are studied. The think-aloud method is well suited for studying users working on complex tasks but requires conditions and means that are not often available, particularly in a field study. Moreover, think-aloud protocol and related techniques may disturb the experimental subjects. Recording user actions in log files is rather easy to perform, and as a consequence, very popular in usability research, although interpretation of user reasoning may be more difficult than with other methods.

Most of these techniques have been used by researchers who study spreadsheet users. In one of the first experiment on users creating spreadsheets with Lotus 1-2-3 (Brown & Gould 1987), participants were videotaped and their keystrokes recorded using specific software. Beyond the elementary actions performed by users, video recording allows researchers to identify more precisely time spent for pausing and planning a solution. Ethnographic techniques mostly relying on interviews of spreadsheet users have been adopted by Nardi and Miller (1990) to study how people cooperate in creating and modifying spreadsheets. We used these techniques in the DidaTab project to record a detailed view of the uses of spreadsheets at home and in school by students. The most recent works rely on data-logging techniques, based on add-ons to the spreadsheet software that records most of the user actions in log files. Adler and Nash (2004) developed a server based software tool (TellTable) to control and record control access, management and use of spreadsheets, and applied this tool for assessing students. In order to study professionals and students detecting and correcting errors, Brain and McDaid (2008) developed a VBA plug-in for Excel that records all cell selection and change actions. The main advantage of this non-intrusive technique is that it allows the recording of the actions of a significant number of participants without modifying their normal working environment.

In a learning context, it is not so easy to define in advance all meaningful actions of users. Although some students may repeat already known errors and processes, some of them may interact in new ways, depending on their usual practice with other computer applications. To cope with this open situation, we adopted a more exploratory approach. Our main objective was to trace most of student gestures in order to be able to guess some of the intentions behind the actions. And we wanted also to be able to replay all the actions performed by the users. With these objectives in mind, we choose a technique to record all events and actions appearing on the screen, even those which are not recorded by ordinary data-logging techniques: video recording of the screen.

### 3.2. Screen capture of screen events

Among several solutions that can be adopted to capture all actions appearing on the screen during interaction, we choose Camtasia Studio© by TechSmith. It captures all onscreen activities: mouse move, selection of a cell or of a range of cells, menu drop-down, click on a command in a menu, window move. In addition, the images taken from a webcam showing the user's face can be included as a "vignette" in the capture. The resulting videos may be saved in several formats and the volume of the corresponding files is reduced by means of a specific video codec.

A free Camtasia player is used to replay all recorded screens. Videos show all events that are traces of students' gestures. In particular, the mouse moves on the screen can give some indications of the information that users are processing, similar to eye tracking methods.

Combined with the recording of the spreadsheet files which provide data concerning students' *production*, screen video captures provides useful data for analysing students' *processes*.

### 3.3. Transcription of events

The most important part of the interaction analysis relies on the transcription phase. The idea is to aggregate very elementary actions into more significant events, more suited for the intended level of analysis.

The screen capture allows detecting all visible changes appearing on the screen, whether they are produced by the user using the keyboard or the mouse, or by the computer application or the operating system as a response to an action.

In the particular case of the transcription of interaction occurring during a test on spreadsheet skills, the whole set of visible events is rather limited and may be identified.

An excerpt of such a transcription is given in Table 2. First, all student actions are reported in detail, except for some of them that are combined into one. For instance, "click on A7" followed by "track to V7" is combined into "select range A7:V7". The feedback given by the spreadsheet software is also recorded with the action. For each action or event, a comment giving additional details or suggesting an interpretation is often added. If no action occurs during a significant period of time, 10 seconds for instance, this "event" is also recorded in the transcription. Then, if a sequence of actions seems to be performed to reach a particular sub-goal, the sequence is aggregated into one *step*. The duration of the step is calculated from the start time of the first action of each step.

From our own experience, the number of detailed actions may vary form 5 to 30 for transcribing one task, depending on the complexity of the task and the activity of the student.

Specific tools like Transana or ActivityLens (Avouris et al., 2007) may be used to assist researchers in that task.

| Start Time (h:min:sec) | Action / Event | Result of action | Comment | Duration (h:min:sec) | Step |
|---|---|---|---|---|---|
| 0:01:52 | no mouse move | | Reads question? | 0:00:21 | Reading |
| 0:02:13 | click on cell X9 | Text cursor appears in the cell | | 0:00:34 | Typing "6+12+.." stopped |
| 0:02:17 | keystrokes in X9: "=6+12+3+…+13" | | Develops a solution then stops | | |
| 0:02:47 | no mouse move | Text cursor blinks at the end of the formula | Thinks of a solution? | 0:00:16 | Thinking |
| 0:03:03 | delete until char = | | | 0:00:25 | Inserting a formula stopped |
| 0:03:22 | click on the menu close to the sigma button | menu showing current functions drops down | Not enough time to read | | |
| 0:03:23 | .. | .. | .. | .. | .. |

**Table 2:** Transcript excerpt of the screen capture of student P8, task 3 (see Fig. 1)

### 3.4. General characteristics collected by transcription

Four main characteristics orient the transcription process. Examples are taken from task 3 (see table 2 and figure 1).

*Time*

Using the start time of actions, it is possible to calculate the time spent on every task, the duration of every step and also of inactive periods, and observe possible increases or decreases in the rate of actions during the whole test.
For instance, student P8 has spent 7' on task 3.

*Sequence of performed tasks*

The sequence of tasks which are successively studied can show navigation between tasks with forward or backward moves.
For instance, student P8 left task 3 after several attempts lasting 3' 21", solved task 7 in 1'07" then returned to task 3 for 3'46".

*Unsuccessful attempts*

All unsuccessful attempts that leave no traces in the files can be observed.
For instance, when working on task 3, student P8 first typed a formula with literals then tried to use a predefined function, then tried to paste the A7:V7 range into cell X9 and finally wrote a formula adding the 22 cell references.

*Software interface uses*

User preferences in using software interface features can be easily observed.
For instance, student P8 entered the formula in task 3 by typing the + sign and selecting cells with a mouse click; student P2 entered the same formula by typing all characters, including cell references.

## 4. EXPERIMENTS

### 4.1. Tests

Tests are built using the methodological framework of the DidaTab project (Tort & Blondel, 2007). Each test includes a series of tasks distributed among five categories of spreadsheet skills. In each task, students are given a spreadsheet that contains data taken from a very simple real life example. They have to either change or transform the appearance of the given spreadsheet (enlarge a column, bold some cells, sort columns), or make calculations from data, or create new objects like charts or tables.

Tests used in the experiment included 11 tasks, and lasted about one hour. Tasks were distributed over skills categories in the following way:
- Editing Cells and Sheets: two tasks (Task 2 Fig. 2 below);
- Writing formulas: five tasks, that require the student to write formulas with arithmetic operators, relative cells references, absolute cells references, functions (Task 3 Fig. 1), to recopy formulas with absolute cells references (Task 7 Fig. 3);
- Creating graphs: two tasks, one that requires creating a curve graph and the other choosing the suitable graphs for two ranges of values;
- Managing data tables: one task that requires sorting a data table on a single criteria (Task 8 Fig. 4 below);
- Modelling: one task that requires building a series of dates.

**Figure 2:** Task 2, Change data display

**Figure 3:** Task 7, Calculate a tax using absolute reference.

**Figure 4:** Task 8, Simple sort.

As Brown & Gould (1987) and Reinhart and Pillay (2005), we found in a previous study (Tort et al. 2008) that students experience the most difficulties when they have to write formulae. We wanted to know if the observation of the process by which they deal with writing formulae would give us information on such difficulties. That is the reason why half of the tasks focussed on writing formulae.

### 4.2. Procedure and data collection

Tests have been administered to two groups of students from the same senior high school, situated in a small town in a suburb of Paris:
- 31 students in the first year of a *"sciences"* stream of study, aged 16-17 (grade 11, SC group).
- 12 students in the second year of a *"social science and economics"* stream of study, aged 17-18 (grade 12, SE group).

Tests were administered on computers, at school, during school time, by the mathematics teacher, who is the same for both groups. The teacher told us that students of the *"social science and economics"* (SE) group used spreadsheet software during his lessons the year before. Students in the *"sciences"* (SC) group did not used spreadsheet in his lessons, but about three quarters of them attended, the year before, an optional course on data processing in physics and chemistry, where they were introduced to spreadsheets and computer programming. It should be also noticed that because of their good level in mathematics, students in the SC group were expected to have few difficulties in using a spreadsheet (on the link between mathematics and spreadsheet abilities see Reinhart & Pillay (2005) and Haspekian & Bruillard (2007).

The application software was Microsoft Excel© 2003. The interface was absolutely the same for both groups, especially displayed toolbars.

Students had to work directly within a given spreadsheet file. An explanation of the whole procedure was given on the first sheet, and the tasks, numbered 1 to 11, were distributed in the next 11 sheets. Students saved their files at the end of the test. Screens were recorded with Camtasia Studio© TechSmith, during the whole duration of the test.

The 31 students in the *sciences* (SC) group and the 12 students in the *social sciences and economics* (SE) group completed the whole tests without any incident; 43 (31+12) spreadsheet files and 43 screen video records were collected. The duration of records extends from 25 minutes to 42 minutes with a mean of 38 minutes.

### 4.3. Abilities revealed by productions

In Tort & al. (2007), we reported on previous tests administered to four groups of students. We found that all the students succeeded in tasks dealing with superficial manipulations of spreadsheet objects that do not need specific knowledge, especially tasks linked to formatting cells. All results decrease slightly when the tasks require more understanding of spreadsheet objects and functionalities. The same is the case for tasks concerning graphs and tables. Moreover, these tasks are more discriminating among the different groups' results. Lastly, the more discriminating tasks where those which dealt with writing formulas.

The results of the two groups studied in the present paper are significantly better (see Tab. 3). In particular, students in the SC group had better results on tasks dealing with objects specific to spreadsheets: decimal number format, cells copying, functions, choosing a suitable graphic and data sorting. Also, the students in the SE group had good results for entity manipulations, writing simple formulas, and graphics. Nevertheless, they had lower results on formulas with functions, absolute cells references, cells copying and data sorting.

| Sub-Category of skills | SC group | SE group |
|---|---|---|
| *Number of students* | *31* | *12* |
| Formatting cells | 31 | 12 |
| Decimal number format | 27 | 10 |
| Writing simple formulas | 29 | 11 |
| Using cells reference | 29 | 12 |
| Cells copying | 24 | 6 |
| Using basic functions | 11 | 1 |
| Using absolute reference | 10 | 2 |
| Creating a graph | 28 | 12 |
| Choosing a suitable graph | 14 | 5 |
| Sorting out data | 20 | 6 |

**Table 3:** Tests Success – Number of students achieving skills

## 5. FIRST RESULTS

We present in this section a first exploratory study of screen records analysis and give main results we have obtained using this specific methodology. We decided to study more in-depth the same number of students in each group for some comparisons and to keep 12 students of each group (total, N=24).

The next section will be devoted to a discussion about this methodology.

The analysis of screen video records was guided by two major aims.

Firstly, we systematically identified the modes of actions used by students on the software interface. It appears that each group of students had specific preferred modes of action: menu commands vs toolbar buttons, keyboard vs mouse, and the preferred actions depend on the performed tasks. It seems that students reproduced actions they have learned.

Secondly, we paid attention to students who did not know how to perform tasks. We look for similar behaviours observed for different students and different tasks.

### 5.1. Preferred modes of action
*Toolbar buttons or menu commands*

Spreadsheet software offers different ways to activate a command: menu commands, contextual menus, keyboard shortcuts and toolbar buttons. Toolbar buttons activate commands with default settings, depending on the current selection. Menus and keyboard shortcuts open dialog boxes, where users can choose the parameters of the command.

Screen videos show that the two tested groups used different modes of action to change cell format (Task 2) and to sort data (Task 8). Students in the SC group used the menu commands and students in the SE group used toolbar buttons (See Table 4).

| Task | Actions | SC group (12) | SE group (12) |
|---|---|---|---|
| Task 2 | Decrease Decimal button | 0 | 10 |
|  | Format cell menu command | 11 | 0 |
| Task 8 | Sort button | 3 | 10 |
|  | Data/Sort menu command | 6 | 1 |

**Table 4:** Approaches used to activate commands - tasks 2 & 8

Moreover, students who easily solved these tasks worked very quickly: less than 50 seconds to change cell format and less than 2 minutes to sort data. We assume that they all repeat a familiar gesture previously learned.

Students who took time to perform the tasks used the toolbar buttons:
- For task 2, screen videos showed 3 students took from 2 to 6 minutes to find the *Decrease Decimal* button, which was not displayed in the toolbar. Then they were lost because the button changes its place once it has been used.
- For task 8, 11 students made several trials to sort data. All of them use toolbar buttons.

*Writing formula in the bar formula or in the cell*

In spreadsheet software, the Formula bar is used to enter and modify the formula of the selected cell. This bar contains several buttons to cancel, to save the input, to open the library of functions. Another way to put a formula in a cell is to type directly the formula in the cell. According to us, this second method maintains the confusion between the displayed value of the cell and its actual content.

| Task | Actions | SC group (12) | SE group (12) |
|---|---|---|---|
| Task 3 | Entry in the Formula bar | 8 | 4 |
| | Entry in the cell | 4 | 8 |

**Table 5:** Approaches used to enter formulas – Task 3

Screen videos show that half students used the formula bar, most often in the SC group.

*Inserting a function in one click or navigating in the library of functions*

The ∑ button is used to insert a function in few mouse clicks. One click on the button inserts the expression "=SUM()" in the selected cell. Moreover, when the selected cell is at the end of a range, the expression is supplemented by the cell references of the range (for instance "=SUM(A7:V7)"). In task 3, such an approach does not work because the required formula is not situated at the end of the data to be summed.

Another method consists of using the "*Insert a function*" menu command. It opens a dialog box, which helps the user to browse the library of functions, choose a function and fill its arguments. A third method consists in directly typing the formula in the cell or in the formula bar. This last method is only useful for users who know the name and the syntax of the function.

| Task | Actions | SC group (12) | SE group (12) |
|---|---|---|---|
| Task 3 | "sigma" button and mouse selection of cells to be referenced | 3 | 1 |
| | "Insert a function" menu command | 3 | 0 |
| | keyboard input of the entire formula | 3 | 1 |
| | "sigma" button and keyboard input of cell references | 2 | 0 |

**Table 6:** Approaches used to insert a SUM function – Task 3

Screen videos showed that students who succeed in inserting a function use various approaches (See table 6). All these students made several trials before obtaining the final formula. Moreover, 4 of them (not reported in the table) failed in the use of the ∑ button.

          

Screen videos showed that most students have not gained a series of methods to perform usual spreadsheet tasks. We presume that they do not follow a well-known step by step process. More likely, their idea of functions is unclear and they need several trials with the interface to succeed in inserting a function.

*Discussion*

We noticed that students who quickly solved a task seemed to reproduce methods that were familiar to them. In addition, each group favoured a different mode of action. Students in the SE group seemed to have experienced direct manipulation and often searched for solutions using this mode of action. Students in the SC group seemed to be more familiar with menus and dialog boxes. This mode of action (menus and dialog boxes) that allows a more detailed control of action is easier to re-apply to various computer applications. According to what we know about the curriculum, it is likely that the SC group spent more time using computers at school; and that the preferred mode of action of these students is partly induced by the type of computer applications they used.

**5.2. Typical approaches in searching for solutions**

In this second part of the analysis, we try to identify the most frequent ways students took when they did not know how to perform a task. The first results of this exploratory work are given below.

*Systematic menu browse*

In this kind of approach, which we named *systematic menu browse*, all menus, options and buttons are browsed. It seems that students take time to read the caption of every option. And if they try one command without obtaining success, they immediately come back to the last seen menu item or button. It looks as if students were looking for one unique command that solved the task. We assume that these students reasoned that there is necessarily one unique command per task to perform, even if a sequence of actions is required.

*Trials of select-command combinations*

We observed students who selected several different cell ranges before applying a command, by way of a button or a menu item. For instance, in the sort task, some students seemed to know which command to apply but several times, they changed the selected cells or the sort button. These trials show a difficulty for students to choose the correct set of data depending on a particular action. This difficulty is reinforced by the spreadsheet software interface which in some cases suggests using another selection before executing the command.

*Opportunistic discovery by analogy*

In theses cases, students acted using analogy between commands. When working on a task, they "discover" a command that they think applicable to solve a previously uncompleted task and then go backwards to the previous task and try the "new" command. We may assume that this kind of analogy is often suggested by the names given to the options of menus. For instance, the label *function* and the choice *sum* in the dialog box opened by the menu *Data Consolidate* may introduce confusion for students trying to sum a cell range in task 3.

*Time, rhythm and density*

The three various approaches described above (*systematic menu browse, trials of select-command combinations, opportunistic discovery by analogy*) have been observed for several students working on several tasks. All these attempts take time and seldom end with a solution.

Moreover, when looking at the screen captures corresponding to these trial and error steps, one can observe that the mouse is always moving and actions follow one another at a high rhythm. In fact, the density of actions, i.e. the number of action per time unit, is greater in these trial and error steps than in other steps.

Within this population of students, those who succeeded in finding a solution after several attempts presented some peculiar ways of working with tasks:
– They spent less time evaluating a result, probably knowing more precisely what they want to see;
– They spent more time reading the content of dialog boxes and messages boxes before choosing an answer.
– They modified their sheets to fix the errors without deviating from the general purpose of the task.
– They immediately repeated a procedure just found.

## 6. DISCUSSION ON BEHAVIOUR RESOLUTION AND FUTURE WORKS

The analysis of screen captures shows a large amount of detail on the various ways adopted by users when working with spreadsheets. Applied to students, it reveals already known defaults and difficulties but also some particular modes of interaction with spreadsheet software.

A rather common behaviour relies on a student belief that each problem has one solution and that this solution has already been scripted in the application software. Some students work as if they just have to find the right button or the right menu item to perform a task. And this could explain what we named "systematic menu browsing". This behaviour can be linked to a common belief of Google users that you just have to ask Google and Google will give you the answer.

Another common behaviour may be seen as a generalisation of "copy and paste". Some students do not base their approach on a step by step process or a model of what has to be done, but on the recall of a similar problem followed by an adaptation with trials and errors. Some evidence of this behaviour may be found in what we named "opportunistic discovery by analogy".

We guess that other interesting ways may be found that are linked to actual modes of interaction with computer applications of the new generation of students. Observation of diverse populations of students would certainly be helpful but it requires important means and screen capture analysis is time consuming. Anyway, it should be useful to establish a catalogue of the kinds of actions carried out by users using spreadsheet software. This could be linked with their level of understanding of spreadsheet objects required to perform these actions and with their practice with other computer applications.

Exploration of duration, rhythm and density has also to be continued. It seems that when users do absolutely not know how to solve a question, they interact very little with the software; when they perfectly know how to solve a question, they perform very few and well organised actions. Between these two extreme cases, users interact a lot more with the software. The density of actions (number of actions / time) seems to be an interesting indicator to investigate.

The results of the exploratory study presented above can be important for the design of a curriculum on spreadsheets. We think that the methods we developed could also be useful to get a better understanding of the various ways professional users are working. Knowing what kind of behaviour users generally adopt in their professional work, depending on specific situations, can be useful in designing appropriate initial or in-service training.